\documentclass[12pt]{article}
\usepackage{epsf}

\newcommand{\bmat}{\left(\begin{array}}
\newcommand{\emat}{\end{array}\right)}

\def\yzero{\smash{\hbox{$y\kern-4pt\raise1pt\hbox{${}^\circ$}$}}}

\def\beq{\begin{equation}}
\def\eeq{\end{equation}}
\def\beqa{\begin{eqnarray}}
\def\eeqa{\end{eqnarray}}

\def\-{\hphantom{-}}

\def\s2{\frac{1}{\sqrt2}}

\def\beq{\begin{equation}}
\def\eeq{\end{equation}}
\def\beqa{\begin{eqnarray}}
\def\eeqa{\end{eqnarray}}

\def\IF{\relax{\rm I\kern-.18em F}}
\def\II{\relax{\rm I\kern-.18em I}}
\def\IP{\relax{\rm I\kern-.18em P}}
\def\IC{\relax\hbox{\kern.25em$\inbar\kern-.3em{\rm C}$}}
\def\IR{\relax{\rm I\kern-.18em R}}

\def\Dsl{\,\raise.15ex\hbox{/}\mkern-13.5mu D} 
\def\IZ{Z\kern-.4em  Z}

\catcode`\@=11   
\newdimen\@rotdimen
\newbox\@rotbox  

\def\@vspec#1{\special{ps:#1}}
\def\@rotstart#1{\@vspec{gsave currentpoint currentpoint translate
   #1 neg exch neg exch translate}}
\def\@rotfinish{\@vspec{currentpoint grestore moveto}}
\def\@rotr#1{\@rotdimen=\ht#1\advance\@rotdimen by\dp#1%
   \hbox to\@rotdimen{\hskip\ht#1\vbox to\wd#1{\@rotstart{90 rotate}%
   \box#1\vss}\hss}\@rotfinish}
\def\@rotl#1{\@rotdimen=\ht#1\advance\@rotdimen by\dp#1%
   \hbox to\@rotdimen{\vbox to\wd#1{\vskip\wd#1\@rotstart{270 rotate}%
   \box#1\vss}\hss}\@rotfinish}%
\def\@rotu#1{\@rotdimen=\ht#1\advance\@rotdimen by\dp#1%
   \hbox to\wd#1{\hskip\wd#1\vbox to\@rotdimen{\vskip\@rotdimen
   \@rotstart{-1 dup scale}\box#1\vss}\hss}\@rotfinish}%
\def\@rotf#1{\hbox to\wd#1{\hskip\wd#1\@rotstart{-1 1 scale}%
   \box#1\hss}\@rotfinish}%
\def\rotate{\@ifnextchar[{\@rotate}{\@rotate[l]}}
\def\@rotate[#1]#2{\setbox\@rotbox=\hbox{#2}\@nameuse{@rot#1}\@rotbox}

\catcode`\@=12

\topmargin
-1.5cm
\textwidth
15.5cm
\textheight
23.5cm
\oddsidemargin
0.7cm
\evensidemargin
1.2cm

\begin{document}

\makeatletter
\@addtoreset{equation}{section}
\makeatother
\renewcommand{\theequation}{\thesection.\arabic{equation}}
\pagestyle{empty}
\rightline{ IFT-UAM/CSIC-04-37}
\rightline{\tt hep-th/0408036}
\vspace{2.5cm}
\begin{center}
\LARGE{ The Fluxed     MSSM    \\[10mm]}
\large{L.E. Ib\'a\~nez \\[6mm]}
\small{
 Departamento de F\'{\i}sica Te\'orica C-XI
and Instituto de F\'{\i}sica Te\'orica  C-XVI,\\[-0.3em]
Universidad Aut\'onoma de Madrid,
Cantoblanco, 28049 Madrid, Spain \\[2mm] and 
\\[2mm]
CERN, Department of Physics, Theory Division,\\
1211 Geneva 23, Switzerland
\\[19mm]}
\small{\bf Abstract} \\[17mm]
\end{center}

\begin{center}
\begin{minipage}[h]{14.0cm}
Recent developments in string compactifications in the presence 
of antisymmetric field backgrounds suggest a new simple
and predictive 
structure for  soft terms in the MSSM
depending only on two parameters. They give
rise to a positive definite scalar potential, 
a solution to the $\mu $-problem, 
flavor universality and absence of a SUSY-CP problem.

\end{minipage}
\end{center}
\newpage
\setcounter{page}{1}
\pagestyle{plain}
\renewcommand{\thefootnote}{\arabic{footnote}}
\setcounter{footnote}{0}


\section{Introduction}

The Minimal Supersymmetric
Standard Model (MSSM) \cite{mssm}
 keeps on being on the most prominent candidates
for an extension of the SM addressing the hierarchy problem.
Gauge coupling unification is an important success  in this
scheme \cite{gcu} and electroweak symmetry breaking
is naturally induced as a consequence of SUSY breaking
and a large t-quark mass \cite{ir2}. 
The weakest point of the MSSM 
is  the origin and structure
of SUSY-breaking. We know that after the dust settles 
one can parametrize our ignorance in terms of dim$\leq 3$  
SUSY-breaking soft terms like gaugino and scalar masses.
Before SUSY-breaking the globally SUSY scalar potential has
the form
\beqa
V_{SUSY}\  & = & \
\vert \ -\mu\ H_d \  + \ h^{ij}_U Q_iU_j \ \vert ^2 
\ +\  \vert \ -\mu\ H_u \  + \ h^{ij}_D Q_iD_j  \ +\
 h^{ij}_L L_i E_j \vert ^2 \\ \nonumber
 & + & \sum_i \  ( \
\vert h^{ij}_U U_jH_u \ +\   h^{ij}_D D_jH_d  \vert ^2 \ +\
\vert h^{ij}_U Q_jH_u \vert ^2 \ +\
\vert h^{ij}_D Q_jH_d \vert ^2 \ +\\ \nonumber
& + &
\vert h^{ij}_L L_jH_d \vert ^2 \ +\   
 \vert h^{ij}_L E_jH_d \vert ^2 \ \ ) \ +\
\ V_{D-terms}  
\label{susypot}
\eeqa
where $\mu $ is the  SUSY Higgs mass and the Yukawa couplings
are complex matrices in generation space.
$Q$ and $U,D$ are the left and right-handed squarks respectively,
whereas $L,R$ are left- and right-handed leptons.
Upon SUSY breaking the most general form of the 
SUSY-breaking yields  terms
\beqa
L_{g}\ & = & \ 
\frac {1}{2} \sum_{a} M_a\ \lambda_a \lambda_a \ + \ h.c. \\
\nonumber
L_{m^2} \  & = & \ -m_{H_d}^2\vert H_d\vert ^2 
-m_{H_u}^2\vert H_u\vert ^2
-m_{Q_{ij}}^2  Q_iQ_j^* 
-m_{U_{ij}}^2  U_iU_j^*
\\ \nonumber
& - & 
m_{D_{ij}}^2  D_iD_j^*
-m_{L_{ij}}^2  L_iL_j^*
-m_{E_{ij}}^2  E_iE_j^*
\\ \nonumber
L_{A,B}\ & = & \ 
-\ A_{ij}^UQ_iU_jH_u
\ -\ A_{ij}^DQ_iD_jH_d
\ -\ A_{ij}^LL_iE_jH_d \  -\ B\ H_dH_u \ +\ h.c.
\eeqa
It is well known that the most general form of soft terms 
has a variety of problems which include
\begin{itemize}

\item {\bf 1.} Lack of flavor universality (at least for the lightest
generations) may induce too large flavor violating neutral currents 
(FCNC).

\item {\bf 2.} Arbitrary complex phases for the $A,B,\mu $ and gaugino 
masses may  lead to large CP-violating electric dipole
moments \cite{efn}.

\item {\bf 3.}
For arbitrary soft terms the scalar potential is
unbounded below and may lead to $SU(2)\times U(1)$ 
breaking at the unification/string scale and/or 
charge- and color-breaking minima.

\item {\bf 4.}
The $\mu $-problem \cite{kn}. We would like to understand why 
a SUSY mass parameter like $\mu $ turns out to be of the
same order of magnitude as the SUSY-breaking mass terms.

\end{itemize}

The first problem is often solved by hand by postulating 
universal scalar masses and $A_{ij}=h_{ij}A$-parameters at the large scale.
This is what is done for example in the popular mSUGRA scenario
\cite{gm}
in which there are  just five parameters $m,M,A,B,\mu $.
In this scenario SUSY is broken in some hidden sector
of the theory at a scale of order $M_{SB}\simeq 10^{10}$ GeV 
and it is transmitted to the observable sector of 
particle theory by  supergravity interactions. The soft 
terms are then of order $\simeq M_{SB}^2/M_{Planck}
\simeq 10^2$ GeV.
Concerning the second problem, again one can postulate that all 
soft terms are real or else that there is some ( so far not
very well motivated) phase alignment taking place. Finally,
in order to get a stable scalar potential one usually restricts
oneself to certain regions of soft parameter space.
Concerning the $\mu $-problem, a natural
mechanism in the context of $N=$ supergravity was proposed
in ref.\cite{giudmas}.

It is clear that in order to find a satisfactory solution
for these problems we need a theory of supersymmetry breaking.
It would be particularly interesting to have a
well motivated underlying theory in which SUSY-breaking takes 
place naturally and in which all the above problems  
are addressed.

Here we are interested in gravity mediated models 
\cite{gm,hlw}
which naturally appear when combining N=1 supersymmetry with gravitational 
interactions. In this case the scalar potential of the massless chiral 
fields is given by the general expression 
\cite{cfgv} (we are using here Planck mass units)
\beq
V\ =\ e^{K}\left(  g^{i{\bar j}}(D_i W)( {\bar D}_{\bar j}{\bar W})
\ -\ 3|W|^2\right) \ +\ D-term
\label{crem}
\eeq
where the index run over all the chiral fields and K and W are
the Kahler potential and superpotential of the theory. One
also has for the Kahler covariant derivative
\beq
D_iW\ =\ \partial_iW \ +\ W K_i
\label{cov}
\eeq
The general idea is that spontaneous supersymmetry breaking 
takes place in some hidden sector of the theory  so that 
 the gravitino gets a mass
$m_{3/2}=exp(K/2)W$. By taking the limit $M_{Planck}\rightarrow \infty $
while mantaning $m_{3/2}$ fixed one generically obtains bosonic
SUSY-breaking soft terms. The gaugino masses are determined by the first
derivative of the gauge kinetic function, i.e. $f_a$ by
\beq
M_a\ =\ (2Ref_a)^{-1} F^i\partial_if_a
\label{gaugino}
\eeq
where the $F^i$ is the vev of the auxiliary field corresponding
to the chiral multiplet $\phi_i$. The  form  of the so 
obtained soft terms thus depend on the structure of $K,W$ and $f_a$,
as well as on which  chiral (hidden sector) fields are  involved 
in the process of SUSY-breaking ($F^i\not=0$).
Thus within this scheme a theory of soft terms correspond 
to a choice for these functions and a minimization of the potential
leading to SUSY-breaking.

A well motivated underlying theory would be string theory.
From the very early days of string theory phenomenology 
attempts were made
to understand the possible origin of SUSY-breaking soft 
terms \cite{din}.  
It was also soon realized that the Kahler moduli $T_i$,
dilaton $S$ and complex structure $M_i$ fields which appear in
string compactifications are natural candidates to constitute
the hidden sector of the theory. For certain classes of heterotic
compactifications (i.e. Abelian  orbifolds and certain large volume limits
of Calabi-Yau (CY) compactifications) it was possible to compute in
perturbation theory the form of the functions $K,W$ and $f_a$
\cite{quevedo}. 
Two natural sources of SUSY-breaking were put forward: gaugino
condensation and a non-vanishing flux $H_{ijk}$ in compact dimensions
for the two index antisymmetric field $B_{ij}$ appearing in the heterotic
\cite{din}.
However the latter source did not look very promising since those
fluxes are quantized and give rise to too large  SUSY breaking of order 
the Planck scale. Gaugino condensation 
\cite{nilles} may lead to hierarchically small
SUSY-breaking, however specific models had two generic problems:
first, there were to many moduli/dilaton fields to be determined
by the gaugino condensation potential; secondly, the vacuum energy
at the minima of the scalar potentials was in general large and negative,
leading to AdS space.

Another slightly more model independent proposal was made in order
to be able to compute soft terms \cite{il,kl,bim,bimrev}.
 The idea was to assume that the
source of SUSY breaking resides in the auxiliary fields of 
the dilaton/moduli fields, e.g. , $F_S,F_{T_i}$. 
Even without knowing what could be the source of these non-vanishing auxiliary
fields, knowledge of the Kahler potential and gauge kinetic
function in some simple heterotic compactifications allowed 
(under the assumption of a vanishing cosmological constant)
for the computation of soft terms as a function of the 
auxiliary fields. Two limits were particularly simple: in the case in
which the auxiliary field of the overall volume field $T$ dominates
($F_T\not=0$) one gets a 'no-scale structure" \cite{ns} leading to a
leading order vanishing cosmological constant. This looks
like a quite interesting starting point, however in that limit 
no soft terms whatsoever were generated (to leading order)
\cite{bim}.
In the dilaton domination case ($F_S\not= 0 $) one obtains
a set of appealing flavor-independent SUSY-breaking soft terms
\cite{kl,bim}.
However no microscopic source for such $F_S\not= 0 $ was 
found. 

In the last nine  years a number of developments have taken
place which suggests to revisit these problems. First, it has been
realized that Type II and Type I strings offer quite promising
possibilities for the construction of string vacua close to
the structure of the SM. A crucial ingredient in this new
model building are Dp-branes, non-perturbative configurations
in string theory  corresponding to $(p+1)$ dimensional subspaces 
inside the full 10-dimensional theory. The crucial property of
$Dp$-branes is that open strings are forced to have their 
boundaries on them. String excitations of open strings 
on the $Dp$-branes give
rise to massless gauge fields as well as fermions and scalars.
Those fields are then to be identified with  the fields of the SM. 
In fact a number of $D$-brane string configurations have been 
constructed using e.g. $D$-branes at singularities \cite{singu} and/or 
intersecting $D$-branes \cite{inter} 
with a massless spectrum remarkably close 
to the SM 
\footnote{One can argue that the semirealistic perturbative 
heterotic models studied up to now are $S$-dual to orientifold
Type IIB compactifications with $D9$-branes. It is thus not
surprising that considering more general configurations with
different $Dp$-branes lead to a new model-building 
possibilities not previously envisaged in perturbative
heterotic compactifications.}.

A second  new ingredient whose importance has  only recently
 been realized is the  role played by antisymmetric 
field fluxes in generic string compactifications
\cite{flux,gkp,kklt} . The case
of 3-form fluxes in Type IIB CY (orientifold) compactifications
has been studied with particular intensity in the last couple
of years. It was realized in \cite{gkp} that
such kind of fluxes in Type IIB orientifold theories 
give rise to a scalar potential which fixes both the dilaton
and the complex-structure moduli $M_i$.  Furthermore the hope exists that,
when including non-perturbative effects depending on the
volume moduli $T_i$ all the moduli in these compactifications 
could be determined \cite{kklt}.   
This would be  an important 
result since the proliferation of undetermined scalar
moduli vevs has been for many years one of the outstanding
problems of string theory.

In a different development it has been recently shown
\cite{grana,ciu,ggjl,ciu2}
that fluxes of this type give also rise to
SUSY-breaking soft terms on the worldvolume of
$D3$-branes and $D7$-branes. 
In particular it was noted  \cite{ciu,ggjl} 
that  fluxes induce non-vanishing expectation values
for the auxiliary fields of the dilaton $S$ and/or
moduli $T_i$, in this way making contact with
the approach followed in refs.\cite{il,kl,bim,bimrev}
and providing a microscopic explanation for the
vevs of the auxiliary fields. 
 In particular in 
ref.\cite{ciu2} certain classes of soft terms
for matter fields in the worldvolume of D7-branes and 
with potential phenomenological interest have appeared.
They are particularly interesting since, 
unlike other previous attempts to compute soft terms from
string theory, they correspond 
to Type IIB orientifold compactifications which
solve the classical equations of motion.
In the present paper we try to obtain general
patterns of MSSM SUSY-breaking soft terms 
based on those recent results.

\section{A bottom-up motivation }

The results suggested by flux-induced SUSY-breaking may
be also motivated from a bottom-up approach.
The first (FCNC) problem of the MSSM suggest 
 to start with flavor independent
 mass  and trilinear terms for squarks and leptons.
Let us consider now the second of the MSSM problems listed
above which concerns complex phases in soft terms.
In a universal setting complex phases 
may appear from $\mu , B,M$ and $A$ parameters.
Physical phases actually depend on  
 the two linear combinations \cite{mssm}
 \beqa
\phi_1 \ =\ \phi_{\mu } \ +\ \phi_{A} \ - \ \phi_{B} \\ \nonumber
\phi_2 \ =\ \phi_{\mu } \ +\ \phi_{M} \ - \ \phi_{B}
\label{fases}
\eeqa
where $\mu =|\mu |e^{i\phi_\mu }$,
$M =|M |e^{i\phi_M }$, $A =|A|e^{i\phi_A }$ and
$B =|B|e^{i\phi_B }$.
For $\phi_{1,2}$ to vanish one needs to have
\beq
\phi_{A}\ =\ \phi_{M} \ \ ;\ \ \phi_{B}\ = \ \phi_{\mu } \ +\ \phi_{M}
\eeq
The simplest way to achieve
this is having soft terms related by:
\beq
A\ =\  aM \ ;\ B \ =\ b\ M\mu
\eeq
with $a,b$ constant real parameters.

It is remarkable that there is a very simple modification of the 
SUSY scalar potential eq.(\ref{susypot}) which solves the first
three problem listed above. This amounts to 
making the replacements
\footnote{As we will see below, from the $N=1$ supergravity
point of view this replacement will correspond to going 
from the SUSY auxiliary fields  
to the SUGRA ones.}
\beqa
  W_{H_u} &  \ &  \longrightarrow \  W_{H_u} \ -\   a_u\ M^*\ H_u^*   
\\ \nonumber
  W_{H_d}  &  \ &  \longrightarrow \  W_{H_d} \ -\   a_d\ M^*\ H_d^*
\label{cambios}
\eeqa
where $W_i$ indicates derivative with respect to the i-th scalar and
 $a_u,a_d$ are real parameters. The superpotential here includes
the bilinear $-\mu H_uH_d$ as usual. 
Note that by making these replacements one
obtains a positive definite scalar potential with soft terms
\beqa
A_u\ & = & \ -a_u\ M \ ;\ A_d\ =\ A_L \ =\ -a_d\ M \\ \nonumber
m^2_{H_u}\ & = &\ a_u^2 \vert M\vert ^2 \ ;\ 
m^2_{H_d}\ =\ a_d^2 \vert M\vert ^2 \ ;\ m^2_{\tilde f}\ =\ 0 \\ 
\nonumber
B\ & = &\ \ (a_u+a_d)M\mu 
\label{softpos}
\eeqa
where $m^2_{\tilde f}$ are the masses of scalar partners of quarks
and leptons. In addition all phases in soft terms may be
rotated away. 
Particularly simple boundary conditions are 
obtained in the  case with $a_u=a_d=1$. In this situation 
all soft terms are determined by a couple of parameters $M$ and $\mu $:
\beqa
m^2_{H_u} &= &m^2_{H_d}=\vert M\vert ^2 \ ;\ m^2_{\tilde f}\ =\ 0
\\ \nonumber
A\ &=& \ -\ M \ \\ \nonumber
 \ B\ & = & \  2M\mu
\label{simple}
\eeqa
Note that in principle one could a similar substitution for the 
rest of the chiral fields of the MSSM
\beq
V_{SUSY}\ =\ \sum_i \vert \partial_i W \vert ^2 \ \longrightarrow
\ V_{SB}\ =\ \sum_i \vert \partial_i W \ -\ a_{i}M^*\phi_{i}^*\vert ^2
\label{compli}
\eeq
with $i=H_{u},H_{d},{\tilde Q}, {\tilde U}_R, {\tilde D}_R ,
{\tilde L}, {\tilde E}_R $. 
Such a procedure would give
rise to universal  mass terms for all squarks/sleptons and
Higgs fields, as well as trilinears.
 Indeed we will see below that flux-induced
SUSY breaking suggests to make such  universal 
replacement with all $a_i=1$.

\section{Fluxes, D-branes and SUSY breaking soft terms}

The kind of string context that we are going to
work on here is that of Type IIB orientifold 
compactifications. This is one of the simplest contexts
in which in the last few years a number of chiral
models with a particle content close to the SM have 
been constructed
\cite{singu,inter}. In these theories one compactifies
Type IIB strings on a CY manifold (or a toroidal orbifold) 
and further modes out the theory by an order-2
twist which includes the $\Omega$ operation which
corresponds to world-sheet parity. Consistency
of the compactification (RR tadpole cancellation conditions) 
requires the presence of some particular set 
of $Dp$ branes with $p=3,5,7,9$ in the setting. 
Depending on the particular form of the orientifold
operation one type or other of $Dp$-branes are required.
If we want to preserve one unbroken $D=4$ SUSY either
$D3,D7$ or alternatively $D9,D5$ sets may  be added. 
In the case in which
the orientifold operation is just $\Omega$ only $D9$-branes
are required and the result is just a standard compactification
of Type I string theory. Since the latter is known to be 
S-dual to the Heterotic string, the effective actions are
quite similar and the phenomenology also is, so 
one does not expect to obtain results very 
different from those previously found
in heterotic compactifications.
 We will rather focus in the case of
orientifolds in which $D7$-branes (and possibly additional
$D3$-branes) are present. The worldvolume  of $D7$-branes
is 8-dimensional and it is supossed to include Minkowski 
space and a 4-cycle inside the compact CY manifold.
If the position of the D7-branes in the transverse 
dimensions is sitting on a smooth point of the CY,
there appears an $N=4$ Yang-Mills theory in the
effective 4-dimensional Lagrangian. If $D7$-branes
sit on top of some (e.g., orbifold) singularity 
the symmetry is reduced and one may get chiral $N=1$
theories (see e.g. ref.\cite{singu} for a description 
of this type of models) of phenomenological interest
\footnote{ More general  Type IIB compactifications of
this type are more efficiently described in terms of
F-theory compactifications on CY 4-folds \cite{ftheory}.}.

As we mentioned, in the last  few  years 
it has been realized the importance of fluxes of
antisymmetric fields on the  structure of
Type IIB orientifold compactifications
\cite{flux,gkp,kklt}. 
Ten-dimensional Type IIB theory has a couple of
antisymmetric tensors $B_{NM}$ and $A_{NM}$ coming respectively
from the so called NS and RR sectors of the theory. They
can have (quantized)  fluxes $H_{ijk}$, $F_{ijk}$ along the compact
complex dimensions $i,j,k=1,2,3$
\beq
\frac{1}{2\pi\alpha'}\int _{C_3} \ F_3 \in 2\pi{\IZ} \quad ;\quad
\frac{1}{2\pi\alpha'}\int _{C_3} \ H_3 \in 2\pi{\IZ}
\label{flujetes}
\eeq
where $C_3$ is any 3-cycle inside the CY.
In SUSY compactifications it is actually the
complex flux combination
\beq 
G_3\ =\ F_3 \ -\ iSH_3
\label{ge}
\eeq
which naturally appears. Here $S$ is the complex axi-dilaton field.
 As long as the
supergravity equations of motion are obeyed this
is a degree of freedom which is generically there and should be
considered.

It was realized in \cite{gkp} that such type of
fluxes give rise to a scalar potential which fixes the
vev of the dilaton and all complex-structure (shape) moduli.
Specifically, $G_3$ backgrounds  of a certain class
(i.e. imaginary self-dual fluxes, ISD
\footnote{ Imaginary self-dual fluxes
verify $G_{(3)}=-i*_6G_{(3)}$, where $*_6$ 
means Hodge dual in the compact six dimensions.} ) solve the equations of 
motion
with a vanishing c.c. to leading order.
The origin of this dynamics is the generation of
a flux-induced superpotential of the form
\cite{gvw,superp}
\beq
W\ =\  \kappa_{10}^{-2}
\int_{M_6} G_{(3)}\wedge \Omega  \ \
\label{superp} 
\eeq
where $\kappa _{10}^2={1\over 2}(2\pi )^7\alpha '^4$ is the
$D=10$ gravitational constant and
 $\Omega$ the Calabi-Yau holomorphic 3-form
(see e.g. ref.\cite{gkp} for details).
This superpotential depends on the dilaton complex 
field $S$ and the complex structure moduli 
(through $\Omega $) but not on the 
kahler moduli ($T_i$ fields). It can be shown that
upon minimization of the resulting $D=4$ 
scalar potential the dilaton $S$ and complex structure
moduli fields are fixed with a vanishing
c.c. (to leading order in both the string coupling constant
$g_s$ and the inverse string tension $\alpha '$).

Furthermore, in \cite{kklt} it was pointed out that,
when combined with non-perturbative effects on the
gauge couplings (like e.g., gaugino condensation) fluxes
may potentially lead to a determination also of
all the  T-like volume moduli. As we said, if true this would
be important progress, since fixing the   
dilaton moduli and complex structure fields in
string theory has always been one of the most outstanding
problems.

As we mentioned it has also been recently shown
that fluxes of this type give  rise to
SUSY-breaking soft terms on the worldvolume of
$D3$-branes and $D7$-branes.
In particular in ref.\cite{ciu} it was realized that 
certain particular choices of $G_3$ backgrounds 
gives rise to soft terms corresponding to 
the dilaton dominance or modulus dominance limits 
discussed in the heterotic literature \cite{bimrev}. 
This is important since it provides for a microscopic 
explanation of non-vanishing auxiliary fields for 
dilaton and moduli. 
The obtained soft terms are proportional to the 
flux densities $G_{(3)}$ which have a dependence for large 
radius  $G_3 \ = \ f{{\alpha '}\over {R^3}}$, 
with $f$ an $R$-independent constant measuring the amount of quantized
flux. Thus one typically obtains SUSY-breaking terms
of order \cite{ciu}
\beq
m_{soft}\ =\ {{g_s^{1/2}}\over \sqrt{2}}G_{(3)}\ =\ {{fg_s^{1/2}}\over \sqrt{2}}
{{\alpha '}\over {R^3}} \ =\ \ =\ {{f\ M_s^2}\over {M_p}}
\label{msoft}
\eeq
with $M_s$ the string scale and $M_p=M_s^4R^3$ the Planck scale.
Thus a way to get soft terms of order the electroweak scale
is having the string scale at an intermediate scale $M_s\simeq 10^{10}$
GeV. However it would be consistent to have a high string scale
with $M_s\simeq M_p$ if the factor $f$ in eq.(\ref{msoft})
is sufficiently small, i.e., if the local flux in the brane
position is for some reason diluted. 
That is for example the case in the
presence of a large warping supresion $\cite{gkp,kklt}$.
In what follows we will  not deal with these issues but assume that 
the resulting soft terms are of order the electroweak scale,
as phenomenologically required. 

It turns out that 
the kind of fluxes  which solve the Type IIB equations of motion
(i.e., ISD ones) do not lead to any soft terms to leading
order for the fields on the worldvolume of $D3$-branes. 
Thus if we try to embed the MSSM on $D3$-branes we
find no soft-terms at all. 
From the effective field-theory
point of view this happens because the ISD fluxes considered
correspond to 'modulus-dominance' SUSY breaking which has a
no-scale structure leading to no soft terms to leading order
\cite{ns}.

The prospects change completely if one considers the
embedding of the SM inside $D7$-branes. 
As we said, this is a natural thing to do in the context of 
Type IIB F-theory compactifications
\footnote{Upon T-dualities and S-duality the type
of setting we are considering should correspond to
a non-perturbative heterotic background in which the 
SM resides on the heterotic 5-branes rather than
coming from the $SO(32)$ or $E_8$ heterotic
gauge groups.}.
As recently emphasized 
in ref.\cite{ciu2} ISD fluxes do give rise to interesting 
SUSY-breaking soft terms for the fields on $D7$-branes. 
We are not giving any details  here
but we can summarize the results and later we will motivate it
from the effective field theory point of view.

A stack of $D7$-branes gives  rise at low-energies to  charged chiral 
multiplets $\phi _i$ upon KK compactification. 
A large class of those admit a  geometric 
interpretation in the sense that vevs for them parametrize the
position of $D7$ brane in transverse space inside the
CY manifold. What we are going to discuss now refers to
that particular class  of D7-brane charged fields
\footnote{In  many F-theory compactifications all 
charged $D7$ zero modes have this geometric character.
In simpler less generic compactifications
(e.g. toroidal or orbifold orientifolds)
some $D7$ zero modes rather parametrize e.g. values of
continuous Wilson lines. We will refer later to those.}

Two types of ISD $G_{mnp}$ flux densities  
(which we take for simplicity to be constant
over the CY) are particularly relevant. 
The first of them corresponds to $(0,3)$ forms
(e.g., a  tensorial structure $G_{{\bar 1}{\bar 2}{\bar 3}}$
in tori)
and gives rise to SUSY-breaking soft terms. The second 
corresponds to $(2,1)$ forms (e.g., a tensorial   
 structure $G_{12{\bar 3}}$ in the toroidal case)  and does not break 
SUSY but may give rise to a $\mu $-term if the symmetries 
of the CY compactication allow for it (see ref.\cite{ciu,ciu2}).
In ref.\cite{ciu2} (section 5.1) the soft terms induced by these types 
of fluxes where computed in  some simple $D7$-brane settings.
It was found that all the bosonic soft-terms 
arise from  positive definite  contributions to the scalar potential
given by
\beq
V_{flux} \ =\ \sum_i
\vert -M^* \phi_i^* \ 
\ +\ \partial_i W \vert ^2
\label{fluxipot}
\eeq
where $\partial _i$ is the derivative with respect to $\phi_i$ 
and $W$ is the   
superpotential involving  the field $\phi_i $.
Here $M$ is  the gaugino  which is 
given in terms of the fluxes by
\beq
M\ =\  c (G_{(0,3)})^*
\label{gaugflux}
\eeq
where $c= \frac {g_s^{1/2}}{3\sqrt{2}} $, with $g_s$ the string
coupling constant.
In addition, if the chiral field $\phi _i $ is vector-like
a possible SUSY mass term appears given by the flux
\beq
\mu  \ =\ -c (G_{(2,1)})^*
\label{valorcillos}
\eeq
As explained in refs.\cite{ciu,ciu2} (see also \cite{ggjl})
these results may be understood also from the effective 
$N=1$ supergravity point of view. Indeed it may be seen that 
a non-vanishing value for $G_{(0,3)}$
induces a non-vanishing expectation value for $F_T$,
the auxiliary field of the overall modulus field $T$.
A constant superpotential proportional
to $G_{(0,3)}$ is also
induced. Now on D7-branes the gauge kinetic function
is simply given by $f_a=T$ and hence from eq.(\ref{gaugino})
gauginos get a mass proportional to $F_T$ (and hence to 
$G_{(0,3)}$).
The SUSY-breaking terms above may be understood 
as arising from a 'modulus domination' scheme. 
This may sound surprising for readers familiar with 
heterotic compactifications, since in that case 
it is well known that modulus domination leads to
no soft terms at all to leading order.
Indeed that would be the case also in Type IIB 
orientifolds if e.g., one 
considers the charged fields arising from $D3$-branes.
Those do not get any soft terms either from 
a $G_{(0,3)}$ background.
It is the fact that our charged fields are living 
on $D7$ branes (which is the natural situation in
F-theory) that makes the difference.
The fact  that modulus dominance may lead to
 non-trivial soft terms for branes different 
than D9 or D3 already appeared 
in ref.\cite{imr} in which the approach of ref.\cite{bim}
was applied to Dp-brane Type I systems (see \cite{ciu2}
for  a description of its connection with flux-induced
soft terms).

From the $N=1$ supergravity point of view is easy to understand
the appearance of this positive definite SUSY-breaking
scalar potential. It is well known that if only the 
auxiliary field of the overall modulus $T$ is breaking 
SUSY, the negative piece of the scalar potential
eq.(\ref{crem}), $-e^{K}3|W|^2$ is canceled by a positive
contribution coming from the T-field auxiliary field
$|D_T W|^2$ leading to a vanishing vacuum energy, this is
the no-scale structure. On the other hand one 
observes that it remains a positive definite piece
which is uncanceled and is given by
\beq
V\ =\ e^{K}\left(  g^{i{\bar j}}(D_i W)( {\bar D}_{\bar j}{\bar W})
\right)
\label{defpoto}
\eeq
where the sum now only runs over the matter fields 
(the contribution of S vanishes identically
since the fluxes considered do not contribute to 
$F_S$). Now
substituting    $D_iW = \partial_iW  + W K_i$,
normalizing canonically the fields  
and recalling that both the constant superpotential
and gaugino masses are proportional to 
$G_{(0,3)}$, one 
obtains the result (\ref{fluxipot}).
Note that this has the same form as the SUSY
scalar potential for matter fields with the only
difference that one replaces the usual derivative
$\partial _i$ by the covariant Kahler
derivative $D_iW$. This is precisely 
the kind of substitution required from
a bottom-up argumentation in the previous section.

Note that sometimes some $D7$ charged fields may not appear 
in this scalar potential. As we said, this is the case of
D7 chiral fields not corresponding to geometric $D7$-brane moduli.
A particular example appears in toroidal or orbifold 
orientifolds in which some massless D7-scalars parametrize
possible continuous Wilson lines. Those fields have kinetic 
terms which depend only on $T$ and then the standard 
cancellation for soft masses characteristic of no-scale
models takes place. Thus in this toroidal case,
scalars $\phi_1$, $\phi_2$ corresponding to Wilson lines in the 
first and second complex planes remain massless whereas 
$\phi_3$, which parametrizes the D7-position is transverse space
get masses in the form described above (see \cite{ciu2}).
However, in generic CY compactifications at most 
discrete (not continuous) Wilson lines may be added,
so the presence of these type of D7-brane moduli is
ungeneric.

\section{The fluxed MSSM}

Although there is has been important recent progress in
obtaining realistic models from Type II orientifolds
\cite{singu,inter}, most of the examples considered
assume vanishing antisymmetric fluxes.
Some preliminary steps on realistic models
with fluxes have  however been given 
 \cite{fluxreal,ciu,ciu2,marchesano} .
Still, although  the $D7$ brane flux configurations
considered  up to now are very simplified,
it could well be that similar structures could appear 
in  more realistic Type IIB orientifolds or 
F-theory compactifications. In particular, the
fact that the scalar potential including soft terms  
is positive definite and involves the 
scalars parametrising the D7-brane positions 
seems to be a general property of SUSY-breaking
induced by ISD antisymmetric backgrounds.

It seems then worth considering in which way 
such a structure could appear in a theory
including the spectrum and interactions of the MSSM. 
A first simple option is to assume that all the
fields of the MSSM correspond to geometric
$D7$-brane moduli in some F-theory compactification.
We then assume that ISD fluxes of type
$G_{(0,3)}$ are present leading to 
modulus dominated SUSY-breaking. In addition,
if a $G_{(2,1)}$ background is present,
the Higgs multiplets may generically get a
$\mu $-term, as explained above.
Then the full SUSY-breaking scalar potential will have the
form
\beqa
V_{FMSSM}
\ & = & \
\vert -\mu \ H_d \ -\   \ M^*\  { H_u}^* \ +\ \sum_{ij}h^{ij}_U Q_iU_j|^2 \
\\ \nonumber
& + &\
|-\mu\  { H_u} \ -\   \ M^*\ H_d^* \ +\ \sum_{ij}h^{ij}_D Q_iD_j
+\sum_{ij}h^{ij}_L L_iE_j \vert ^2 \\ \nonumber
& + \sum_i  &  \ ( \ 
\vert  -\ M^*{Q_L^i}^*\ +\  h^{ij}_U U_jH_u \ +
\   h^{ij}_D D_jH_d  \vert ^2 \ \\ \nonumber 
& + & \
\vert          -\ M^*{U^i}^*\ +\ h^{ij}_U Q_jH_u \vert ^2 \ +\
\vert   -\ M^*{D^i}^*\ + \ +  h^{ij}_D Q_jH_d \vert ^2 \ +\\ \nonumber
& + &
\vert  -\ M^*{E^i}^*\ + \   h^{ij}_L L_jH_d \vert ^2 \ +\
 \vert  -\ M^*{L^i}^*\ + \ h^{ij}_L E_jH_d \vert ^2 \ \ ) \  ) \  +\
\ V_{D-terms}
\label{auxi}
\eeqa
This potential gives rise to the following set of
bosonic SUSY-breaking soft terms for the MSSM: 
\beqa
m^2_{H_u} &= &m^2_{H_d}= m^2_{\tilde q}=m^2_{\tilde l}=
             \vert M\vert ^2 \\ \nonumber 
A_U\ &=& \  A_D \ =\ A_L \ = \ -\ 3 M \ \\ \nonumber
 \ B\ & = & \   2M\mu
\label{simplest}
\eeqa
Note that all soft terms are universal and given by 
only two free parameters $M$ and $\mu $ which are
determined by the ISD fluxes  
$G_{(0,3)}$ and
$G_{(2,1)}$ respectively. 
As a general remark note that, since both the 
SUSY-breaking parameter $M$ and the $\mu-term$ arise
from fluxes, it is natural for them to be of the same 
order of magnitude. Thus flux SUSY-breaking solves
naturally the $\mu $-problem.

We thus see that under the assumption that 1) 
our SM fields are embedded as geometric D7-brane 
fields in a general Type IIB orientifold (or more generally, F-theory)
compactification and 2) that ISD fluxes are present 
we obtain a rather simple structure of soft terms addressing
the four MSSM problems listed at the beginning of the paper
\footnote{In the addendum we address a possible generalization which includes
the case in which the MSSM particles live at the intersections of
different stacks of D7-branes.}.

Most of the features of the above simplest  choice of soft terms
may be also obtained from a simple  $N=1$
supergravity toy model. Indeed,
consider the following string motivated type of
gauge kinetic function $f_a$ and Kahler potential
\beqa
f_a \ & = & \ T \\ \nonumber
K \ &=&\ -log(S+S^*\ -\ |H_u+H_d^*|^2-\ \sum_i |\phi_i|^2) \ -\
 3log(T+T^*) \\ \nonumber
W(S) \  &= &\ a M_{p}^2\ S \ +\ b M_{p}^2
\label{kahler}
\eeqa
where $\phi_i$ represent the squark and slepton fields.
The superpotential $W(S)$ is modeling the general
flux-induced superpotential eq.(\ref{superp}) and
$a,b$ are  complex constants  related  to the flux densities 
$H_{(3)},F_{(3)}$ integrated over the CY space
\footnote{Note that this flux inspired
superpotential has the form of a simple
Polony superpotential for the dilaton
complex field. However the 'Polony field' here is
$S$ which does not have a canonical metric as in
the old supergravity models.}. 
We will however treat $a,b$ as 
free constant parameters.
Readers familiar with string derived $N=1$ models
will note a number of differences from the standard
perturbative heterotic lore. First, the gauge kinetic
functions are not given by the axi-dilaton field $S$ 
but rather by the complexified volume field $T$.
A second difference is that the MSSM fields appear
in the Kahler potential  
combined with the complex dilaton $S$ field rather than
the overall modulus $T$.
 However this is precisely the form
of the gauge kinetic functions and Kahler potential   
which appear when considering simple
toroidal/orbifold  compactifications of
Type IIB orientifolds with 
geometric (no-Wilson-line) D7-brane matter fields 
   \cite{imr,lmrs} .
In this toy example is easy to see that
upon minimization of the scalar potential 
one has $F_S=0$, which fixes the  value of the 
complex dilaton field at $S=(b^*/a^*)$.
At the minimum a constant superpotential is
obtained at $W_0=M_p^2[(a/a^*)b^*+b]$. 
Due to the no-scale structure of the $T$-field 
SUSY is broken by  a non-vanishing 
$F_T$, with a vanishing vacuum energy,
leaving the T-vev undetermined.
Using standard supergravity formulae (see e.g. ref.{\cite{bimrev})
it is an easy exercise to show that precisely the
simple choice of soft terms (4.2) are
obtained. This particular form of Kahler potential leads
to a contribution to the
$\mu$-term  generated
a la Giudice-Masiero $\mu _0 =M$
\cite{giudmas} . More generally
the flux analysis shows  that the complete $\mu $-term and
gaugino masses $M$ depend on different fluxes and
hence are in general independent parameters.
Thus one can reproduce this more general case by considering
an explicit $\mu $-term in the original superpotential.

Irrespective of its string theory motivation, the
above simple choice of Kahler potential, gauge  kinetic 
term and $S$-dependent superpotential constitute
an interesting $N=1$ supergravity model containing
the MSSM spectrum. No fine-tuning is required to
get a (tree-level) vanishing cosmological constant,
still it gives rise to universal SUSY-breaking 
soft terms.

Let us end with a number of comments. The above simple
universal result is obtained under the (reasonable) assumption that 
all the MSSM fields are D7-geometric moduli. We have mentioned,
however that in certain cases some D7 zero modes do not have
a geometric meaning but rather correspond to, e.g., the
possible existence of continuous W.L. backgrounds on the
D7-brane worldvolume. Those may remain massless even 
in the presence of fluxes.  
 Thus one may perhaps 
consider other ways of embedding the MSSM inside 
Type IIB orientifolds in which some of the MSSM
chiral fields do not get masses. That may lead to 
non-universal scenarios in which some of the MSSM fields
get soft masses and others don't. These mixed scenarios 
seem however less natural that the universal one described above,
since, as we mentioned, the presence of continuous Wilson
lines is ungeneric in CY compactifications. 
Also, from the phenomenological point of view 
many of those non-universal possibilities are problematic
due to FCNC constraints. One possibility which would be
quite simple and 
universal would be one in which 
only the Higgs multiplets correspond to D7 geometric moduli.
In this case the only source of scalar SUSY-breaking
soft terms would be those arising from 
the first two terms in eq.(\ref{auxi}) which lead to
 soft terms of the form 
$m^2_{H_u} = m^2_{H_d}=\vert M\vert ^2\ $   ;$\ m^2_{\tilde f}\ =\ 0$;
$A_U =  A_D  = A_L \  =  -\ M$ , $ 
  B\  =  \  \ 2M\mu $.
There is finally a third possibility also leading to universal
soft terms, which is to assume that all chiral fermions 
correspond to D7 geometric moduli, but not the Higgs fields.
In that case the obtained soft terms have $m^2_{\tilde f}=|M|^2$,
$A_U=A_D=A_L=-2M$ and $m^2_{H_u}=m^2_{H_d}= \mu ^2 = B =0$.  
Note that soft terms like these two  may be obtained from the
toy supergravity model above by having the
squark and slepton fields combined with the 
T-field(S-field)  in the no-scale fashion in eq.(\ref{kahler})
while mantaning the Higgs(squark/slepton)  fields combined with $S$
respectively. Let us however emphasize again that these other
non-universal possibilities look less generic in the context of 
F-theory.

A further comment concerns gaugino masses and gauge coupling unification.
If the relevant D7-branes containing the MSSM fields have
all the same geometry (i.e., wrap the same 4-cycle in the CY compact 
space) and are located at the same point in transverse space,
gauge coupling unification at the string scale is expected.
In that case there will also be in general a universal 
gaugino mass parameter. This has been our simplifying assumption
above, although generalizations without this property could 
 be envisaged.

\section{Final comments}

We have argued  that the presence  of antisymmetric 
fluxes in Type IIB orientifolds 
with D7-branes (or, in general, F-theory
compactifications in complex 4-folds) give rise to
an interesting class of soft SUSY-breaking soft terms,
eq.(4.2). 
An important point to emphasize is that the
relevant class of fluxes studied (imaginary self-dual
3-form fluxes) solves the classical equations of
motion for compactified Type IIB string theory \cite{gkp}. Thus the class 
of models discussed gives  rise, to leading order, to
consistent $N=1$ low-energy theories with softly
broken soft terms. To our knowledge, this is the first 
class of classical string compactifications in which
that is the case.

From the low-energy $N=1$
supergravity effective Lagrangian point of view the
presence of fluxes give rise to  modulus dominance
SUSY-breaking. However, unlike similar type
of SUSY breaking studied in perturbative heterotic
compactifications (or its Type I dual with D9 branes),
in the present case interesting classes of soft terms do
appear on the massless fields coming from D7-branes
after compactification. 

Assuming that the MSSM fields may be embedded as
geometric D7-brane moduli, we have argued that the 
relevant MSSM soft terms depend only on two
free parameters, the gaugino mass $M$ and a $\mu $-term.
Those are in turn given by certain classes of 
flux densities. Thus one expects both parameters
to be of the same order of magnitude, given the fact that
they have a common origin, fluxes.
The SUSY-breaking scalar potential
is positive definite and it is simply obtained from
the SUSY scalar potential by making the
replacement
$\partial_iW\rightarrow D_iW$, with i running over all
the chiral multiplets of the MSSM.
The set of soft terms (4.2) so obtained is universal
and solves the SUSY-CP problem due to the 
specific relationships obtained between the
$A,B,M$ and $\mu $ parameters. 

A natural question to ask is whether one could find 
a scheme in which $M$ and $\mu $ were related. 
In that case  we would be left with a single 
parameter describing all soft terms. 
In principle the flux densities 
$G_{(0,3)}$ and $G_{(2,1)}$
are independent parameters. In fact, in generic CY
compactifications there are a number of different 3-cycles 
through which fluxes can exist. 
The integral of the 
corresponding RR and NS 3-forms over 3-cycles 
in the CY are quantized (eq.\ref{flujetes}) although
the flux densities themselves are not. Thus parametrically
these densities 
$G_{(0,3)}$, $G_{(2,1)}$
 go like $\simeq N/(Vol_C)$, with N integers 
and $Vol_C$ the volume of the corresponding 3-cycle.
In specific compactifications the volumes of the
different cycles could be related (e.g. equal) and
one would expect specific relationships 
(e.g. $\mu = 2M $) depending on the different integers.
To find this kind of relationships would require however
an specific example of compactification yielding 
the MSSM.

A number of other interesting issues should be addressed. 
Of course, it would be important to have orientifolds
or F-theory examples with a massless spectrum 
as close as possible to that of the MSSM. Also
it would be important to realize specific examples
(perhaps with some large dimension transverse
to D7-branes  and/or warping or other) in which
the size of soft terms is 
naturally of order the electroweak scale.
We have also ignored in our analysis the
dynamics which eventually determines the 
volume moduli $T$ (see ref.\cite{nilles} for a recent
discussion of this issue). 
Another obvious point is to study the low-energy
spectrum of SUSY masses as well as the generation of
radiative electroweak symmetry breaking in this
class of theories. The latter will be presented elsewhere
\cite{bi}.

\vspace*{0.5cm}

\vspace*{0.5cm}

\section{Addendum on Intersecting $D7$-branes}

Some of the most promising semirealistic Type IIA string
orientifold models are based on intersecting D6-branes
\cite{inter}. Upon T-duality this class of models may 
be equivalently described as Type IIB orientifold models
with intersecting D7-branes with magnetic fluxes in their
worldvolume  (branes of other dimensions may
also be present in particular models). Thus it would be
interesting to see what kind of soft terms are induced
by ISD fluxes in this type of intersecting models. 
Most of those models are toroidal (or orbifold) compactifications.
 So we will   consider here a case with a toroidal
compactification on $T^2\times T^2\times T^2$ 
in which we 
have three classes of D7-branes, $D7_i$,  $i=1,2,3$ which
are transverse to the i-th 2-torus respectively. 
Thus in addition to the fields coming from the worldvolume
of a stack of D7-branes $D7_i$ considered in the main text, 
we will now have in general new chiral fields $\phi_{ij}$
corresponding to open strings living at the intersections of 
a pair of distinct stacks of branes $D7_i-D7_j$. Let us study here
what kind of soft terms appear for these extra fields living at the
intersections in the presence of ISD fluxes (corresponding in field theory
language to modulus domination). 
 Using the effective Lagrangian approach 
for the overall modulus dominance in
refs.\cite{imr,lmrs} as well as \cite{ciu2,ggjl} one can figure
out what to expect. One finds that the bosonic SUSY-breaking 
soft terms for all chiral fields  may be obtained as arising from a 
slight generalization of the results in the main text, namely
\beq
V_{SUSY}\ =\ \sum_i \vert \partial_i W \vert ^2 \ \longrightarrow
\ V_{SB}\ =\ \sum_i \ (1-\xi_{i}) \vert \partial_i W \ -\ M^*\phi_{i}^*\vert ^2
\ +\ \sum_i \xi_i \vert \partial_i W \vert ^2
\label{complimas}
\eeq
where the T-dependence of the metric of the field $\phi_i$
is $(T+T^*)^{-\xi_i}$
\footnote{ This may be considered as the Type IIB analogue
of the modular weights of ref.\cite{il,bim}).}. 
Thus the case considered in the previous sections 
(geometric D7-brane moduli, no T-dependence in the
kinetic term of the field) corresponds to $\xi_i=0$ 
and the case of Wilson-line D7-fields (no S-dependence in the
metric) would correspond to 
$\xi_i=1$, giving rise to no soft terms. From section 7 in ref.\cite{imr}
(see also \cite{ggjl})
one can see that  the matter fields in $D7_i-D7_j$ intersections
have metric proportional to $((S+S^*)(T+T^*))^{-1/2}$ and 
then one has $\xi_i=1/2$ for the fields. One can easyly check that
eq.(\ref{complimas}) with $\xi_i = 1/2$ indeed reproduces the
bosonic soft terms in \cite{imr}.
It is easy to understand qualitatively these results. For fields with
metric proportional to $(T+T^*)^{-1}$ (corresponding to $\xi=1$) 
the usual no-scale cancellation gives zero soft terms for the fields.
For fields with metric proportional to $(S+S^*)^{-1}$
there is no cancellation at all and soft terms appear as in previous sections.
On the other hand the metric for the fields at intersections
is in some sense in between and there is a partial no-scale cancellation
of soft terms, giving rise to the $\xi_i=1/2$ factor.
Note that in more general intersecting D7-brane models in which
there are magnetic fluxes in their worldvolume the T-dependence 
of the metric of the fields at intersections will in general depend
on the (magnetic) fluxes so that the $\xi_i$ will be model dependent
and have  to be computed in each model.

Let us now apply these ideas to the MSSM. In order to have 
universality the $\xi_i$'s corresponding to different generations of the 
same quark or lepton should be equal. But in fact in the class of
models under discussion that is in general the case, different 
generations of fields have the same metric. Thus universality is
a natural situation even in these more general configurations.
So appart from the flux-induced  parameters $M$ and $\mu$, there
will be a set of 7 model dependent parameters 
$\xi_i$, $i=Q,U,D,L,E,H_u,H_d$. In terms of all these and using
eq.(\ref{complimas}) one can write down the following set of
soft terms arising from ISD fluxes for the MSSM:
\beqa
m_i^2\ & = &\ (1-\xi_i)|M|^2 \ \ , \ \ i=Q,U,D,L,E,H_u,H_d \ \\ \nonumber
A_U\  & = & \ -M(3-\xi_{H_u}-\xi_Q-\xi_U) \\ \nonumber 
A_D\  & = & \ -M(3-\xi_{H_d}-\xi_Q-\xi_D) \\ \nonumber
A_L\  & = & \ -M(3-\xi_{H_d}-\xi_L-\xi_E) \\ \nonumber
B \ & = & \ M\mu (2\ -\ \xi_{H_u}-\xi_{H_d})
\label{cojosoftgen}
\eeqa
It must be emphasized that  in a given model
the $\xi_i$ are computable quantities determined by the 
T-dependence of the metric of the corresponding field and
hence only  $M,\mu$ remain as free parameters.
Note that for all $\xi=0$ we recover the situation described  
in the previous sections, eq.(\ref{simplest}).
As in the simpler case discussed in the previous sections, 
these soft terms may be derived from a simple 
$N=1$ supergravity model with the same gauge kinetic 
function and S-dependent superpotential as in eq.(\ref{kahler})
but with a generalized Kahler potential 
in which the metric of matter fields include the 
mentioned $(T+T^*)^{-\xi_i}$ dependence (see \cite{imr,kn,lmrs}).

It would be interesting to compute the ISD flux-induced soft 
terms in specific semirealistic compactifications. 
An example of this is the  Type II intersecting D-brane
configuration yielding an MSSM-like spectrum proposed in
\cite{guay}. This local D-brane configuration 
may be embedded into a full $N=1$ SUSY Type IIB orientifold \cite{csu}
$Z_2\times Z_2$ compactification \cite{marchesano} with
additional ISD fluxes. In the latter constructions
the MSSM fields appear at the intersections of 3 sets of
branes $D7_i$, $i=a,b,c$, very much as described above.
A stack of 8 D7-branes $D7_a$ give rise to the 
Pati-Salam group $SU(4)$ (which may easily be broken down to
$SU(3)\times U(1)_{B-L}$ in the presence of Wilson line
backgrounds). Two parallel D7-branes $D7_b$($D7_c$) give
rise to the gauge group   $SU(2)_L$($SU(2)_R$).
The $D7_a$ stack 
has in addition magnetic flux in its worldvolume, giving rise to 
the replication of generations. It is beyond the scope of the 
present paper to give a detailed description of the soft terms
induced in a model like this (see refs.\cite{lrs2,fi} fo recent
analysis).
It may be however ilustrative to figure out simple main features
of the expected structure of soft terms.
In this particular model the stacks $D7_b$ and $D7_c$ have
no magnetic flux in their worldvolume. Thus the fields at their intersection
(one set of MSSM Higgs fields) will simply have $\xi_{H_u}=\xi_{H_d}=1/2$.
The quarks and leptons reside at intersections $D7_a-D7_b$ and 
$D7_a-D7_c$. Given the symmetries  of the  brane configuration
in this model (and the built-in Pati-Salam symmetry) all
quarks and leptons are universal, $\xi_Q=\xi_U=\xi_D=\xi_L=\xi_E=\xi$.
So all in all the general form of soft terms will be
\beqa
m_{H_u}^2& = & m_{H_d}^2\  =  \frac {|M|^2}{2} \\ \nonumber 
m_Q^2&=&m_U^2=m_D^2=m_L^2=m_E^2= \ (1-\xi)|M|^2 \  \\ \nonumber
A_U  & = &  A_D=A_L\ =\  -M(5/2-2\xi) \\ \nonumber
B \ & = & \ M\mu  \ \ \ .
\label{softguay}
\eeqa
For large T-values the magnetic fluxes are diluted and 
one expects to recover the case without fluxes with
$\xi \simeq 1/2$. In that limit one would have the universal result
\beqa
m_{i}^2& = & \frac {|M|^2}{2} \ \ ;\ \ 
i=Q,U,D,L,E,H_u,H_d \ \\ \nonumber
A_U  & = &  A_D=A_L\ =\  -M(3/2) \\ \nonumber  
B \ & = & \ M\mu  \ .
\label{softguay}
\eeqa

\medskip

\vspace*{0.5cm}

\vspace*{0.5cm}

\vspace*{0.5cm}

\medskip

{\bf \large Acknowledgments}

I  thank B. Allanach, R. Barbieri,  
A. Brignole, P.G.C\'amara, F. Marchesano,  
F. Quevedo, G. Ross and A. Uranga  for 
useful  discussions.
I am also grateful  to  CERN's Theory Division for hospitality.
This work has been partially supported by the European Commission under
the  RTN contract HPRN-CT-2000-00148 and the CICYT (Spain).

\end{document}